\documentclass[pdflatex,sn-mathphys-num]{sn-jnl}

\usepackage{graphicx}%
\usepackage{multirow}%
\usepackage{amsmath,amssymb,amsfonts}%
\usepackage{amsthm}%
\usepackage{mathrsfs}%
\usepackage[title]{appendix}%
\usepackage{xcolor}%
\usepackage{textcomp}%
\usepackage{manyfoot}%
\usepackage{booktabs}%
\usepackage{algorithm}%
\usepackage{algorithmicx}%
\usepackage{algpseudocode}%
\usepackage{listings}%
\usepackage{array}
\newcolumntype{P}[1]{>{\centering\arraybackslash}p{#1}}
\usepackage{longtable}
\usepackage{changepage}
\usepackage{comment}
\usepackage{natbib}

\begin{document}
\title{Alternate Groundwater Modelling Strategies: A Multi-Faceted Data-Driven Approach}

\author*[1]{\fnm{Muralidharan} \sur{Kunnummal}}\email{muralikustat@gmail.com}\equalcont{These authors contributed equally to this work.}

\author*[1]{\fnm{Agniva} \sur{Das}}\email{agniva.d-statphd@msubaroda.ac.in}
\equalcont{These authors contributed equally to this work.}

\author[1]{\fnm{Shrey} \sur{Pandya}}\email{shrey.pandya-stat@msubaroda.ac.in}

\author[2]{\fnm{Jong-Min} \sur{Kim}}\email{jongmink@morris.umn.edu}

\affil*[1]{\orgdiv{Department of Statistics}, \orgname{Faculty of Science, The Maharaja Sayajirao University of Baroda}, \orgaddress{\city{Vadodara}, \postcode{390002}, \state{Gujarat}, \country{India}}}

\affil[2]{\orgdiv{Statistics Discipline, Division of Science and Mathematics}, \orgname{University of Minnesota-Morris}, \orgaddress{\city{Morris}, \postcode{MN56267}, \state{Minnesota}, \country{United States of America}}}


\abstract{The impact of statistical methodologies on studying groundwater has been significant in the last several decades, due to cheaper computational abilities and presence of technologies that enable us to extract and measure more and more data. This paper focuses on the validation of statistical methodologies that are in practice and continue to be at the earliest disposal of the researcher, demonstrating how traditional time-series models and modern neural networks may be a viable option to analyse and make viable forecasts from data commonly available in this domain, and suggesting a copula-based strategy to obtain directional dependencies of groundwater level, spatially. This paper also proposes a sphere of model validation, seldom addressed in this domain: the model longevity or the model shelf-life. Use of such validation techniques not only ensure lower computational cost while maintaining reasonably high accuracy, but also, in some cases, ensure robust predictions or forecasts, and assist in comparing multiple models.}
\maketitle
\section{Introduction}\label{sec1}

Groundwater is globally acknowledged as one of the sources of water stock used to deal with water scarcity. The volume of withdrawal isn’t reflective of its social value. Local availability of groundwater ensures the potential of regulating withdrawal on demand, drought aversion, and lower purification costs, thereby ensuring a higher economic value as compared to surface water. Therefore, identification of groundwater recharge zones is critical for sustainably managing groundwater resources. \par
Groundwater Recharge (GWR) refers to the water moving from an unsaturated area into a saturated area beneath the water table surface, which includes the associated flow away from the water table within the saturated zone. When water moves past the groundwater level and infiltrates into the saturated zone, recharge is said to have occurred. Its determination conventionally relies on the quantity of precipitation and the percolation process. Groundwater accounts for 98\% of the unfrozen freshwater on earth, according to the International Groundwater Resources Assessment Centre (IGRAC) (operating under UNESCO). Groundwater is responsible for al-most half of the world's drinking water, around 40\% of irrigation water, and about one-third of the water used by industry. The IGRAC performs various functions worldwide, ranging from promoting sustainable groundwater management, developing new groundwater resource maps, to enhancing groundwater governance policies, and safeguarding water user rights in emergencies. In addition to sustaining ecosystems, groundwater is also an essential element in adapting to climate change.\par
Looking at it from an Indian standpoint, the extent to which groundwater is extracted varies throughout the country. According to the Economic Survey of India (2021-2022), the percentage of units categorized as "safe" (Stage of Extraction or SoE $< 70\%$) has decreased from $73\%$ in 2009 to $64\%$ in 2020. The percentage of "semi-critical" (SoE $> 70\%$ and $\le 90\%$) units has increased from $9\%$ in 2009 to $15\%$ in 2020. The proportion of "critical" (SoE $> 90\%$ and $\le 100\%$) units has remained between $3\%$ and $5\%$ from 2004 to 2020. During 2004-2020, the proportion of "over-exploited" (SoE $> 100\%$) units accounted for $14\%$ to $17\%$ of the total assessment units. Moreover, nearly one percent of the assessment units have been classified as "saline". It has also been observed that overexploitation of ground water resources, i.e., extraction exceeding the annually replenishable GWR is concentrated in north-west and parts of southern India.\par 
For many decades, the quality of most of the aggregated data that was used by national as well as international agencies that had been taken from national and international databases, was extremely difficult to assess and considered infeasible \cite{Kircher1665, zektser2004groundwater}. Data quality varied heavily, and still does in the present day and age, owing to large differences in data acquisition practices from country to country – even inside countries – and the varying percentage of areas for which no real data but only rough estimates are available. This provides motivation for the development and adoption of mathematical, statistical, and subsequently, machine learning models, coupled with various deep learning frameworks that act as tools for mapping groundwater dynamics. \par 
In recent times, the use of Geographic Information System (GIS) data, combined with conventional maps and on-the-ground data, has created an easement in establishing the baseline information for groundwater potential zones, and as such, several conventional graphical and statistical tools and techniques have been utilized by researchers to analyze and interpret groundwater quality \cite{Karanth1987, Sara1991, Fawcett1994, guler2002evaluation, Machiwal2010}. However, modern techniques such as time series modelling, multivariate statistics, and geostatistical modelling, among others, are now being applied to precisely characterize and better understand groundwater quality for efficient management and protection of groundwater resources \cite{guler2002evaluation, Jha2007, Cloutier2008, Steube2009, Machiwal2010, Machiwal2015}. These techniques help differentiate between anthropogenic and natural processes or factors influencing groundwater quality. \par 
In regions that are arid or semi-arid, where groundwater is the primary source of freshwater due to the lack of surface water scarcity, groundwater resources are often overallocated, particularly in coastal areas \cite{carrion2021geometric}. Groundwater salinization and quality degradation are common in these regions, and these issues may be caused by natural processes such as water-rock interaction, marine intrusion, or the impact of saline adjacent aquifers. Anthropogenic activities like over-pumping, domestic or industrial sewage leakage, or groundwater pollution from intensive agricultural practices may also contribute to groundwater salinization \cite{Shiri2013, An2018, Machiwal2018, Amiri2021, Bahir2021}. \par 
Groundwater level variation is highly nonlinear and depends on various factors such as precipitation, evapotranspiration, soil characteristics, and watershed topography \cite{Khedri2020}, which complicate the task of groundwater level prediction. The literature presents two primary approaches to groundwater level prediction: data-driven or statistical models and numerical models. Classical numerical models like MODFLOW, FEFLOW, and GMS (Groundwater Modelling System) require expensive input data that may not always be available \cite{Movahedian2016}. Data-driven models, on the other hand, require fewer assumptions about the underlying mechanisms used in numerical models and have proven more useful in evaluating uncertainty, vagueness, variability, and complexities inherent in water resources and environmental management problems \cite{Nayak2006, Mohanty2015}. Classical numerical methods are often more uncertain due to data inaccuracies, high data requirements and costs, and limited hydrogeological parameters used in groundwater simulation and geological heterogeneity within a model grid. \par
The primary focus of study is to extract as much information on the groundwater dynamics in a region, with minimal data. The information that we choose to focus on, however, pertains more to a “surveillance-friendly” environment, i.e. data that is expected to be open source, fairly regularly available, and can be captured in real-time. It may be noted that most of the groundwater monitoring systems available today do not collect data in real-time, however, extensive research in remote sensing and GIS may render this possible in the near future. The frameworks discussed in the paper can be used in both scenarios equally well. However, increased data availability may enhance its forecasting and pattern recognition abilities. \par 
This paper aims to introduce alternative models and techniques in statistical literature that might be helpful in navigating some of the aspects of groundwater level estimation that have been under-explored. We aim to serve two crucial purposes in a groundwater level estimation study:
\begin{enumerate}
    \item To quantify the extent of effect of variables like temperature and precipitation have on the groundwater level at a location (in our case, it is that of the Patyapura Station in Vadodara, Gujarat). However, we are also interested in exploring the causality patterns over time, i.e.
    \begin{enumerate}
        \item We explore the direction of the effect thus quantified to gage the causality of a particular variable on the groundwater level.
        \item We explore the effect of a one-sigma shock on each of the variables determine the extent to which this shock induces variations in the other variables.
        \item We also explore how this dependence pattern varies over time.
        \item Forecast future groundwater level values based on the available data. Additionally, we aim to compare our model's forecasting ability with modern deep models (in our case, the vanilla Long Short Term Memory Model).
        \item Assess the quality of forecast based on model longevity, i.e. the length of time that a model can forecast accurately without having to “train” on additional data.
    \end{enumerate}
    \item To assess the dependence of groundwater level at one spatial point (in our case, a station/well) on all other spatial points from which data has been collected (only one variable: Groundwater Level), to capture dependence patterns between stations/regions.
\end{enumerate}
Additionally, we shall discuss some of the notable methodologies and highlight alternative strategies for more efficient and sustainable models.

\section{Literature Review}\label{sec2}

By convention, a groundwater model refers to a scale or electric model of an aquifer representative of natural groundwater flow. They may include several aspects of the groundwater, like quality or chemical composition, specifically aimed at predicting movement and aftermath of the chemical in diverse scenarios, or predicting the effects due to variation in the behaviour of aquifer(s). Majority of the computations pertaining to groundwater models are expressed through groundwater flow equations, characterized by differential equations solved using various numerical (approximation) methods, alternatively known as mathematical, numerical, or computational groundwater models. \par 
A plethora of statistical and machine learning models have been in use to ascertain groundwater potential and locate groundwater recharge zones for many years, sometimes hybridized as per requirement, and optimized thereafter \cite{Magesh2012, Kattimani2018, Khedri2020, Azma2021}. 
\subsection{Groundwater Level Modelling}
The authors \cite{Rajaee2019}, in their review of artificial intelligence-based models for groundwater level modelling (GWL), have already covered much ground on several models that are in use today. In their paper, they have reviewed models like ANN’s (Artificial Neural Networks) \cite{Lallahem2005, Daliakopoulos2005, Nayak2006, Feng2008, Krishna2008, Nourani2008, Tsanis2008, Banerjee2009, Yang2009, Beale2010, Chen2010, Mohanty2010, Sreekanth2011, Trichakis2011, Rakhshandehroo2012, Taormina2012, Sahoo2013, Ying2014, Gholami2015, Juan2015, Sun2016, Kouziokas2018, Wunsch2018, Lee2019}, ANFIS (Adaptive Neuro-Fuzzy Inference System) \cite{Jalalkamali2011, Shirmohammadi2013, Emamgholizadeh2014, Khaki2015, Mirzavand2015, Gong2016}, Genetic Programming models \cite{Orouji2013, Shiri2013}, SVM’s (Support Vector Machines) \cite{Yoon2016, Huang2017, Nie2017, Mukherjee2018, Guzman2019, Tang2019}and some hybrid AI techniques including various combinations of classical AI methods, mentioned above, with different efficient methods for input data pre-processing, including wavelet analysis \cite{Adamowski2011, Kisi2012, Maheswaran2012, Moosavi2013, Moosavi2014, He2014, Suryanarayana2014, Khalil2015, Nourani2015, Yang2015, Nourani2016, Wen2016, Barzegar2017, Zare2018}, Kriging \cite{Nourani2011, Tapoglou2014}, Self-Organizing-Map (SOM) \cite{Chang2016, Han2016} based clustering, ant colony optimization (ACO) \cite{Hosseini2016} and Genetic Algorithm (GA) \cite{Jha2015} optimization. However, they have covered most of the methodologies up until 2017. Since then, there has been an influx of literature, particularly in that sphere. \par
The authors of \cite{Natarajan2020} compared the performances of various AI algorithms in predicting GWL at six different locations in the district of Vizianagaram, Andhra Pradesh, India. The data comprised of monthly GWL observations from six locations, namely Cheepurupalli, Gantyada, Garugubilli, Jiyyammavalasa, Pachipenta and Vizianagaram for the period of 2007–2012. Techniques used include Artificial Neural Network (ANN), Genetic Programming (GP), Support Vector Machine (SVM), Extreme Learning Machine (ELM), and some SVM-hybrids like the SVM-QPSO (Support Vector Machine – Quantum-behaved Particle Swarm Optimizer) and SVM-RBF (Support Vector Machine – Radial Basis Function). ELM was found significantly better for prediction of Ground Water Level (GWL), as compared to the other techniques, and the SVM-QPSO hybrid ran at par with the ELM on certain occasions, coming a very close second. Model performance was evaluated based on the Root Mean Squared Error (RMSE), Pearson’s Correlation Coefficient, Coefficient of Determination ($R^2$), the Mean Absolute Error (MAE). Note that the Pearson’s Correlation Coefficients and the $R^2$ values were obtained for the combinations of observed and predicted values for each location, as opposed to the residuals versus fitted/predicted values. It is also noteworthy that all performance metrics pertain to training samples only, i.e., no out of sample predictions were made by the models concerned, as neither a training-testing holdout split, nor any sort of cross-validation was done. \par 
The authors \cite{Banadkooki2020} conducted a study on predicting groundwater level (GWL) using precipitation and temperature data with various temporal delays. They utilized three models: radial basis function (RBF) neural network with whale algorithm (WA), multilayer perception (MLP) with WA, and genetic programming (GP) to predict GWL. To assess the uncertainty related to the input parameters of the models, the authors considered three scenarios with different inputs. In scenario one, the input data consisted only of the average temperature of the region and three temporal delays of 3, 6, and 9 months were considered. In the second scenario, the authors evaluated the predicted GWL based on the precipitation data of 3, 6, and 9 months. In the third scenario, they simultaneously used data from average precipitation and temperature. The authors concluded that the first scenario produced the best results.\par
A recent paper by Zhu et al.\cite{zhu2025robust} presents a probabilistic groundwater level forecasting framework using a multi-model machine learning ensemble integrated using Bayesian Optimization Algorithm (BOA) and a stacking algorithm by using Copula functions to construct joint distributions of the loss function, only to obtain hyperparameter estimates using maximum likelihood method, and subsequently obtaining posterior probability distribution functions for uncertainty analysis. The study identifies previous groundwater depth, meteorological factors (rainfall, evaporation), hydrological factors (runoff), and crop water demand as input variables. Rule-based constraints are applied to optimize lag selection. Five models:
\begin{itemize}
    \item Random Forest (RF)
    \item Support Vector Machine (SVM)
    \item Gradient Boosting Decision Tree (GBDT)
    \item eXtreme Gradient Boosting (XGBoost)
    \item Artificial Neural Networks (ANN) (Multi-Layer Perceptron)
\end{itemize}
are trained for groundwater depth prediction. Bayesian optimization is employed to fine-tune hyperparameters, and a stacking algorithm is used to combine individual models for improved accuracy. The study fits nine candidate probability distributions: 
\begin{itemize}
    \item Normal
    \item Gamma
    \item Log-normal
    \item Pearson - III
    \item Laplace
    \item Gumbel
    \item Cauchy
    \item Generalized Extreme Value
    \item Generalized Pareto
\end{itemize} 
to the observed groundwater data and selects the best-fit distribution using Kolmogorov-Smirnov (K-S) tests, Sum of Squares Error (SSE), and Akaike Information Criterion (AIC). Five Copula functions:
\begin{itemize}
    \item Gaussian
    \item t-Copula
    \item Gumbel
    \item Clayton
    \item Frank
\end{itemize} 
are tested to model the dependence between observed and predicted groundwater depths, with the best function selected based on Kolmogorov-Smirnov test and RMSE.
The study evaluates model accuracy using MAE, RMSE, Pearson Correlation Coefficient(R), and Nash-Sutcliffe Efficiency (NSE).

\subsection{Groundwater Recharge Zone Classification Models}
In recent years, the challenges associated with sustainable groundwater storage management have compelled researchers worldwide to develop Groundwater Potential across varied geological zones. Many studies, in that regard have tried using statistical models to get further optimized results. \par 
Among many methodologies that are in regular use, notable among them has been the one proposed by \cite{Jaafarzadeh2021}, aimed at identifying potential groundwater recharge (GWR) zones using a classifier ensemble, by combining a Maximum Entropy and Frequency Ratio model (variant of Decision Tree Classifiers), with validation criteria in the form of Correctly Classified Instances (CCI) (confusion matrix approach) and area under the Receiver's Operating Characteristic (ROC) Curve indices. They had collected data from Marboreh Watershed located between $33^{\circ}12^{\prime} N$, to $33^{\circ}03^{\prime} N$ and $49^{\circ}03^{\prime} E$ to $49^{\circ}57^{\prime} E$ in the west of Iran, Lorestan Province. As assumed by the authors, the dependence of GWR on permeability and soil’s absorption aims to justify double-ring infiltrometer method and soil sampling methodologies adopted for model performance evaluation as well as for understanding the spatial variability of GWR. A 70:30 holdout split was done on the percolation points for model training and testing. 15 effective factors on GWR potential were used in this study:
\begin{itemize}
    \item Elevation (TIF) (generated from ASTER DEM data)
    \item Slope Percentage (TIF) (calculated from slope degree) (generated from ASTER DEM data)
    \item Slope Aspect (TIF) (calculated from slope angle) (generated from ASTER DEM data)
    \item Profile Curvature (TIF)
    \item Plan Curvature (TIF)
    \item Drainage Density (HIF) (generated from ASTER DEM data)
    \item Distance from rivers (HIF)
    \item Topographic Wetness Index (TWI) (HIF) (generated from ASTER DEM data)
    \item Stream Power Index (SPI) (HIF) (generated from ASTER DEM data)
    \item Rainfall (HIF)
    \item Lithology (GIF)
    \item Fault Distance (GIF)
    \item Land Use (EIF)
    \item Normalized Difference Vegetation Index (NDVI) (EIF)
    \item Soil Texture 
\end{itemize}
	
These 15 factors were then classified into topographic influencing factors (TIF), hydrological influencing factors (HIF), geological influencing factors (GIF) and ecological influencing factors (EIF). Apart from those mentioned, rest of the factors were extracted from Landsat – 8 images and through conventional soil and geology data. These factors were then mapped using ArcGIS. \par 
In a semi-arid region of Iran, a spatial assessment of the groundwater potential in an aquifer was carried out by \cite{Yariyan2020}. The study utilized geographic information systems (GIS)-based statistical modelling and sampled water from 75 agricultural wells across the Marvdasht Plain, measuring the electrical conductivity (EC) of the water samples.  The data consisted of broadly three categories: Topographic variables included Altitude, Slope, Slope Aspect, Curvature, Topographic Ruggedness Index (TRI), Topographic Wetness Index (TWI); Earthly variables include Lithology, Soil, Distance to Fault, Distance to Stream; Hydrological variables include Rainfall, Drainage Density, Flow Direction. Only the Land Use variable was collected using Landsat – 8 OLI Images. To detect multicollinearity present in the dataset, Variance Inflation Factor (VIF) and tolerance were opted for. Various combinations of different regression models, including the Frequency Ratio (FR) model and the Evidential Belief Function (EBF) model, were hybridized with the Radial Basis Function (RBF) NN, Index of Entropy (IoE) model, Fuzzy Art Mapping (FAM) NN. For model validation, a 70-30 Holdout was used and then, the authors opted for the ROC approach, i.e. metrics focused on included the Area Under Curve (AUC), Positive Predictive Value (PPV), Negative Predictive Value (NPV), Sensitivity, Specificity, Accuracy. Friedman's Test was also used for model validation. \par 
\cite{razavi2019groundwater} attempted to introduce a new ensemble data mining approach with bivariate statistical models, using Frequency Ratio (FR), Certainty Factor (CF), Evidential Belief Function (EBF), Random Forest and a Logistic Model Tree. The study area was restricted to the Booshehr plain, between the latitudes $27^{\circ}50^{\prime}N$ and $28^{\circ}30^{\prime}N$, and longitudes $51^{\circ}20^{\prime}E$ and $52^{\circ}10^{\prime}E$ in south eastern Iran. This paper takes a 70:30 holdout split over the data from multiple sources; information about groundwater wells collected from government sources and topographic data from ASTER DEM downloaded in a spatial resolution of 30m $\times$ 30m, their layers created through ArcGIS and SAGA GIS software(s). Factors considered for modelling included altitude, slope angle, slope aspect, plan curvature, profile curvature, topographic wetness index (TWI), slope length, distance to fault, fault density, distance to river, drainage density, land use, soil, lithology, rainfall. All these factors were grouped into the following group of parameters: 
\begin{itemize}
    \item \textbf{Topographic:} Altitude, profile curvature, \item slope length, slope angle, plan curvature, and slope aspect.
    \item \textbf{Hydrological:} Distance to river, TWI (calculated using slope angle and cumulative upslope), drainage density, distance to river.
    \item \textbf{Geological:} Lithology, distance to fault, fault density.  
    \item  \textbf{Climate:} Rainfall. 
    \item \textbf{Ecological:} Land use, soil.
\end{itemize}
The paper uses several isolated ensembles of combinations of Frequency Ratio, Certainty Factor and Evidential Belief Function models paired with two types of ensemble structures namely, Random Forests and Logistic Model Trees. Model was validated using the Receiver's Operating Characteristic (ROC) curve and the Area Under the Curve (AUC) metric associated with it.
\subsection{Artificial Neural Network Models for Groundwater}
Artificial intelligence (AI) methods have been widely used in various water re-sources applications since their inception. Groundwater level prediction studies have used a combination of AI methods such as artificial neural network (ANN), fuzzy logic (FL), adaptive neuro fuzzy inference system (ANFIS), group method of data handling (GMDH), and least squares support vector machine (LSSVM) \cite{Shirmohammadi2013, Suryanarayana2014, Mohanty2015}. ANNs have been used for groundwater level prediction \cite{coulibaly2001artificial} while ANFIS has attracted significant interest in groundwater level modelling \cite{Shirmohammadi2013, Emamgholizadeh2014, Nourani2016}. Although there are few applications of GMDH for modelling environmental systems \cite{Samsudin2010}, several researchers have reported the successful application of the LSSVM approach for groundwater level prediction \cite{Yoon2016, Lee2019}. Comparisons between the performance of ANN models in predicting groundwater models based solely on hydrological data have been carried out \cite{coulibaly2001artificial}, as well as evaluations of several types of ANNs and training algorithms in predicting monthly groundwater level fluctuations in an aquifer in the Messara Valley, Crete \cite{Daliakopoulos2005}. \par
A hybrid AI meshless model has been developed for spatiotemporal groundwater level modelling \cite{Nourani2016}. Using this model, the time series of groundwater levels observed in different piezometers were first de-noised using a threshold-based wavelet method, and the effect of de-noised and noisy data was compared in temporal groundwater level modelling using ANN and ANFIS. The ANFIS RBF model was found to be more reliable than the ANN-RBF model in both calibration and validation steps. The authors of the paper cited in \cite{Gill2007} compared the performance of ANN with support vector machine (SVM) in groundwater-level prediction under missing or incomplete data conditions, and the potential of the SVM model was generally found to be superior to the ANN model for input combinations. \cite{Yoon2016, Lee2019} developed two nonlinear time-series models for predicting the groundwater level fluctuations using ANNs and SVMs, showing that SVM was better than ANN during both model training and testing phases. Similarly, \cite{Emamgholizadeh2014} employed ANN and ANFIS to predict the groundwater level of Bastam Plain in Iran, and both models were found to predict groundwater level accurately. \par 
Recent studies have compared the predictive capacity of several AI techniques in forecasting groundwater level. Shiri et al. \cite{Shiri2013} compared the abilities of Gene Expression Programming (GEP), ANFIS, ANN, and SVM methods in groundwater level fore-casting for up to 7-day prediction intervals, demonstrating that the GEP models were better in forecasting water table level fluctuations up to 7 days beyond data records. \cite{Shirmohammadi2013} employed several data-driven techniques, such as system identification, time series, and ANFIS models to predict groundwater levels for various forecasting periods. The results indicated that ANFIS provided accurate predictions of groundwater levels for 1- and 2-month lead times. Although AI data-driven models have seen an increase in hydrologic researches on rainfall \cite{Nastos2014, Lian2020}, streamflow \cite{Shu2008, Samsudin2010, Sanikhani2012, Tiu2018, Azma2021}, and drought \cite{Soh2018, Fung2020}, few studies have compared widely accepted models (e.g. ANN, FL, ANFIS, and LSSVM) for groundwater level prediction in the literature \cite{Shiri2013, Emamgholizadeh2014}.\par
More recently, the study by Weng et al. \cite{weng2025groundwater} proposed a novel wavelet-deep learning approach to predict groundwater levels in the Daliao area of Kaohsiung, Taiwan, integrating real-time IoT pumping data. The methodology combines wavelet transform for feature extraction with deep learning models, including Recurrent Neural Networks (RNNs), Long Short-Term Memory (LSTM) networks, and Gated Recurrent Units (GRUs). They have utilized hourly data on:
\begin{enumerate}
    \item GWL
    \item Rainfall
    \item Pumping Data
    \item Temperature
    \item Tidal Information
    \item Humidity
\end{enumerate} 
from August 23, 2017, to January 30, 2020, collected from various sources, including Taiwan's Central Weather Administration (CWA) and the Water Resources Agency (WRA). Three types of signal analysis was done to execute the feature extraction process:
\begin{itemize}
    \item \textbf{Continuous wavelet transformation (CWT) analysis:} Represented as a wavelet magnitude diagram or a Scalogram,
    \begin{equation*}
        S_{c_x}(a,b)=\left| x_\omega(a,b)\right|^2
    \end{equation*}
    where, for some continuous time signal $x(t)$,
    \begin{equation*}
        x_\omega(a,b)=\frac{1}{\sqrt{|b|}} \left| \int_{-\infty}^\infty {x(t) \ \Psi\left(\frac{t-a}{b} \right) dt} \right|^2, 
    \end{equation*}
    $\Psi(t)$ denotes the parent wavelet function, continuous with respect to time and frequency, $a$ is the position of translation and $b$ is the magnification factor. Generally used to decompose a continuous time signal.
    \item \textbf{Inverse continuous wavelet transformation (ICWT) analysis:} Used as a filter to reconstruct the components decomposed by CWT (only at a specified frequency),
    \begin{equation*}
        x(t)=\int_0^\infty \int_{-\infty}^\infty {\frac{1}{b^2} \ x_\omega(a,b) \ \frac{1}{\sqrt{|b|}} {\Psi_1\left(\frac{t-a}{b} \right) \ da} \ db}
    \end{equation*}
    \item \textbf{Cross-Wavelet Transform:} Used to identify causal relationships between two time series $x_t$ and $y_t$. It is expressed as
    \begin{align*}
        w_t^{xy}(s)&=w_t^x(s) w_t^y(s) \\
        &=\left| w_t^{xy}(s)\right| e^{t \varphi(s)}
    \end{align*}
    where, $w_t^x$ and $w_t^y$ are the respective time series $x_t$ and $y_t$ post wavelet transform at frequency $s$, $w_t^{xy}(s)$ denoting the cross-wavelet spectogram, $\left| w_t^{xy}(s)\right|$ denoting the power post crossed wavelet, and $\varphi(s)$ denotes the phase angle betweem $x_t$ and $y_t$ in the frequency domain. This was later used to compute wavelet coherence and time-delay between the two-time series. In our study, we used the Granger-Causality measure to achieve the same.
\end{itemize}
The authors of \cite{weng2025groundwater} compared the performance of three deep learning models:
\begin{itemize}
    \item A traditional Recurrent Neural Network (RNN) Model
    \item Long Short-Term Memory (LSTM) Model 
    \item Gated Recurrent Unit (GRU) Model
\end{itemize}
to forecast GWL, through the features extracted using the above wavelet analyses. This course of action was bifurcated further along the lines of incorporation of IoT information (pumping data).  
The wavelet analysis in \cite{weng2025groundwater} was executed using the MATLAB wavelet toolbox. Model performances were measured using traditional statistical indicators and metrics like the $R^2$, MAE, MSE (\& RMSE) and AIC. The findings in this study, suggest that including smart IoT-based pumping data and wavelet-transformed time-frequency analysis significantly enhances groundwater level predictions. However, no model validation techniques have been employed to underline the reproducibility of model output.\par
Sun et al.\cite{sun2024deep} apply a Transformer-based deep learning model to predict groundwater levels in the Zhuoshui River basin.The dataset includes 725 ten-day interval records from 2000–2019, collected from 8 groundwater stations, 20 rainfall stations, and 4 flow stations. The Transformer model is compared against three benchmark models: (1) MLP, (2) Convolutional Neural Network (CNN), and (3) LSTM. The Transformer model incorporates self-attention and multi-head attention mechanisms to capture long-term dependencies and relationships between groundwater, rainfall, and flow levels. For model evaluation, the authors used $R^2$ and MAE.The Transformer model outperformed CNN, LSTM, and MLP in most cases, achieving the highest R² values and lowest MAE scores. It was found to be most effective at capturing groundwater fluctuations, though some stations exhibited lower accuracy due to over-extraction or geological variability. However, the following points must be considered as well:
\begin{itemize}
    \item The study compares different neural network models, but it does not include traditional statistical time-series models (e.g., ARIMA, Kalman Filters) as baselines. These could have provided insight into whether deep learning significantly outperforms simpler models.
    \item While $R^2$ and MAE are useful, no confidence intervals or statistical significance tests are presented to determine whether the Transformer’s superiority over CNN and LSTM is statistically significant.
    \item The study does not mention cross-validation or bootstrapping to assess model generalizability, which could lead to overfitting concerns.
    \item The paper does not clearly state whether the input variables were checked for multicollinearity. Since rainfall, river flow, and groundwater levels are highly interdependent, correlated inputs might reduce model interpretability.
    \item The $Z$-score transformation of groundwater levels is appropriate for normalization but might have benefited from additional preprocessing methods, such as detrending or seasonal decomposition, especially since hydrological data exhibits periodicity.
    \item The study assumes a fixed learning rate (0.0001 for Transformer), but it does not discuss hyperparameter tuning techniques (e.g., grid search, Bayesian optimization), which could have optimized performance further.
    \item The study does not perform feature importance analysis (e.g., SHAP values) to determine which variables contribute most to predictions. While the Transformer’s self-attention mechanism is theoretically beneficial, its practical interpretability in groundwater hydrology remains unexplored.

\end{itemize}

\section{Research Methodology} \label{sec3}
\begin{figure}[ht]
    \centering
    \includegraphics[width=1\linewidth]{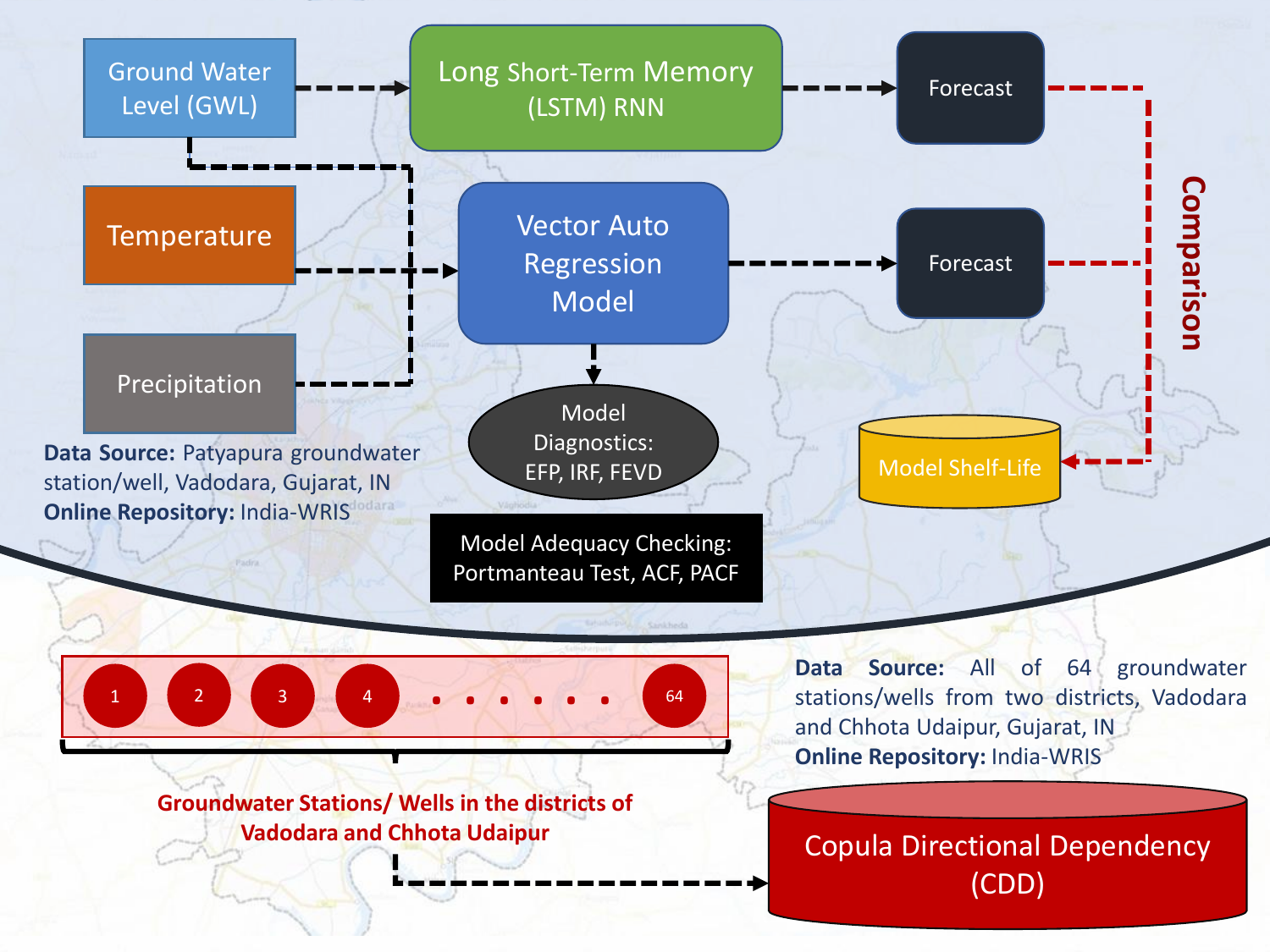}
    \caption{A pictoral representation of research outline of this article, including information about variables in the data, its source, the methodologies adopted for the analysis and the chronology of the model building process.}
    \label{fig:GWM}
\end{figure}

As discussed at length in Section \ref{sec1}, we aim to provide an alternative framework for analysing and interpreting groundwater level data. A visual representation is provided in Figure \ref{fig:GWM}.

\begin{enumerate}
    \item For the first objective, we use the Vector Auto Regression (VAR) model which is a powerful statistical tool used to capture the interdependencies among multiple time series variables. In this case, we have utilized it in understanding the dynamic relationships between temperature, precipitation, and groundwater levels, as these variables are inherently interconnected and evolve over time. However, these interactions are often lagged and nonlinear, which makes the use of simpler time-series models redundant, thereby establishing the VAR model as a well-suited framework for analyzing such systems. In the VAR framework, all variables are treated as endogenous, allowing the model to simultaneously account for the mutual feedback between temperature, precipitation, and groundwater levels. For example, the model can capture how a change in precipitation today affects groundwater levels in subsequent months while also considering how past groundwater levels might influence precipitation patterns through climatic feedbacks. This makes the VAR model particularly effective in describing the temporal dynamics of the system. Furthermore, it helps quantify causal relationships and predict future groundwater levels based on historical temperature and precipitation patterns, aiding in water resource management.
    \begin{enumerate}
        \item As a component of the VAR framework, the Impulse Response Function (IRF) assesses how a shock to one variable (e.g., a sudden increase in temperature) propagates through the system, impacting groundwater levels and precipitation over time.
        \item Another component of this framework, i.e. the Granger-Causality is a statistical concept used to determine whether one time series can predict another. In the context of a Vector Auto Regression (VAR) model, it examines whether the past values of one variable (e.g., precipitation) provide significant information about the future values of another variable (e.g., groundwater levels), beyond what is already explained by the latter's past values. It serves to provide more insight into the direction of causality between the variables, how it evolves over time and determines time-lags with significant causality, thereby providing insights into the temporal scale of these interdependencies. For example, if precipitation “Granger-causes” (read. causes) groundwater levels, this indicates that past precipitation data can help predict future groundwater levels. Similarly, if temperature “Granger-causes” precipitation, it suggests that climatic feedbacks are influencing rainfall patterns. This information can guide water resource planning and climate adaptation strategies. 
    \end{enumerate}
    \item The Copula Directional Dependence (CDD) measure provides a flexible tool for analyzing directional dependencies in environmental data, accommodating non-normal distributions and nonlinear relationships. This approach can be instrumental in understanding the spread and control of air pollution across different regions, aiding policymakers in designing targeted interventions. We propose the CDD measure, utilizing beta regression models to assess directional relationships without relying on strict assumptions of specific probability distributions or linearity. By introducing the CDD measure, this study offers a robust framework for assessing directional dependencies in groundwater level data among groundwater wells/stations. The findings underscore the dynamic nature of inter-city groundwater relationships and highlight the importance of considering directional dependencies in environmental studies.

\end{enumerate}
The Vector Auto Regression (VAR) model offers distinct advantages for analyzing the dynamic relationships between temperature, precipitation, and groundwater levels, making it a superior choice compared to other popular statistical methodologies for certain scenarios.
Firstly, unlike univariate models (e.g., ARIMA), VAR simultaneously models multiple interdependent time series, allowing for the exploration of feedback loops among temperature, precipitation, and groundwater levels. Secondly, many regression-based methods, like linear regression, require a clear distinction between dependent and independent variables. VAR assumes no such hierarchy, enabling it to capture bidirectional influences, such as how groundwater levels might influence precipitation patterns. VAR, also, explicitly incorporates lagged values of all variables, making it well-suited for studying time-delayed effects, such as how precipitation impacts groundwater recharge over weeks or months, provides a robust framework for testing Granger causality, helping identify the directional and temporal causative relationships among variables, a feature not readily available in many other techniques, and finally, its ability to model interdependencies makes it a reliable tool for forecasting future values of all variables in the system, a capability often lacking in simpler models like ordinary regression. \par 
In comparison with other methodologies, VAR fares exceptionally well. While linear regression can establish relationships between variables, it does not account for temporal dependencies or feedback loops, which are crucial in climate-groundwater systems. Auto-Regressive Integrated Moving Average (ARIMA) focuses on a single variable's temporal dynamics and lacks the ability to model multivariate interdependencies, making it less appropriate for systems where variables interact. Machine learning techniques like random forests or neural networks are powerful but often act as "black boxes," lacking the interpretability of VAR. Additionally, they may require large datasets and are prone to overfitting in small-sample contexts common in environmental studies. Structural Equation Model (SEM) can model multivariate relationships but assumes predefined causal pathways, whereas VAR allows the data to determine the directionality and strength of interactions. Therefore, it is safe to claim that the VAR model strikes a balance between simplicity, interpretability, and analytical power. Its ability to capture dynamic, lagged interactions among interdependent variables makes it particularly suited for studying climate-groundwater systems. While other methodologies may excel in specific contexts, VAR’s versatility and focus on temporal dynamics make it an excellent choice for this analysis.\par 
Monitoring groundwater levels across multiple wells or stations often involves complex interdependencies influenced by spatial, climatic, and hydrological factors. In our study, however, we aim to use as little information as possible to explore spatial dependencies only.  Traditional methods for analyzing such relationships may struggle to capture nonlinear dependencies or directional influences between stations. The Copula Directional Dependence (CDD) measure offers a robust framework for overcoming these challenges.\par 
The motivation for using CDD lies in its ability to quantify directional relationships between groundwater wells without assuming linearity or specific distribution forms. Groundwater levels at one well can influence or be influenced by others through subsurface water flow, recharge patterns, or anthropogenic activities such as pumping. By applying CDD, it becomes possible to model these dependencies while accounting for variations due to geography, aquifer properties, and climatic inputs like precipitation, and several other factors. However, as our focus is more on trying to explore these dependencies on spatial data, solely, and due to a severe dearth of related data for all 64 stations in Vadodara city, we shall not be accounting other factors in our model. \par 
Furthermore, CDD’s flexibility in handling nonlinear relationships and its capability to analyze lagged effects make it particularly suitable for identifying cause-effect dynamics in groundwater systems. For example, it can reveal how changes in one well propagate to others over time or, in case one were to incorporate other factors in the model, how external factors, such as heavy rainfall at a recharge zone, influence downstream wells. This directional understanding is crucial for sustainable water resource management, enabling targeted interventions and predictive modelling. As such, the CDD measure provides a powerful tool for capturing the spatial and temporal dependencies among groundwater wells, making it an innovative approach to enhance groundwater monitoring and management strategies.
\subsection{Dataset}
For the illustration of model shelf–life estimation, we have considered the problem of groundwater level prediction/forecasting for a single station, namely Patiyapura (22.275$^{\circ}\text{N}$, 73.44167$^{\circ}\text{E}$), in Vadodara, India, based solely on monthly temperature and precipitation, collected over four quarters of 20 years from 2000 to 2020 and the first quarter of 2021 (85 time points), partitioned into a 70 – 30 Holdout (59 points in training set; 26 points in testing set). This data was sourced from \url{https://indiawris.gov.in/wris/#/groundWater}. Run charts for the quarterly observed groundwater level, temperature and precipitation levels are provided in the latter section. \par
For obtaining the copula directional dependencies, we have a separate dataset (source: \url{https://indiawris.gov.in/wris/#/groundWater}) consisting of groundwater levels of 64 groundwater stations in Vadodara, Gujarat, India, collected over four quarters of 20 years from 2000 to 2020 and the first quarter of 2021. \par

\subsection{The Vector Autoregression model}
For forecasting groundwater level, we have used the simple vector autoregression model (VAR), a multivariate time series prediction/forecasting model that is extensively used for analysing economic and financial data. However, from a statistical standpoint, one may choose to use it as a model for groundwater level prediction/forecasting, as it is basically a natural extension of the univariate autoregressive model to dynamic multivariate time series. It can be done with a simultaneous equation model, as the study considered several endogenous variables in a single model. In VAR analysis, there is usually no exogenous (independent) variable in the model.\par 
Let ${{X}}_t=\left[X_{1t},\ X_{2t},\ \ldots,\ X_{nt}\right]^\prime$ denote an ($n\times1$) vector of time series variables. The standard form of the basic $p$-lag VAR system is of the form:
\begin{equation}
    {{X}}_t={c}+{\Gamma}_{1}{{X}}_{t-1}+{\Gamma}_{2}{{X}}_{t-2}+\ldots+{\Gamma}_{p}{{X}}_{t-p}+{{\epsilon}}_t\ ,\ \ \ t=1,\ 2,\ \ldots,T
\end{equation}
where,  ${\Gamma}_{i}$ are ($n\times n$) coefficient matrices, and ${{\epsilon}}_t$ is an $\left(n\times1\right)$ unobservable zero mean white noise which is serially uncorrelated (independent), with time invariant covariance matrix ${\Sigma}$. \par 
Lag length $p$ is conventionally determined using various model selection criteria, like the Akaike Information Criterion (AIC), Bayesian (Schwarz) Information Criterion (BIC/SIC), Hannan-Quinn Information Criterion (HQIC), Final Prediction Error (FPE), all of which were consulted and resulted at $p=4$, unanimously. Also, the ACF and PACF plots had induced a similar suggestion.\par 
To address the assumption of zero serial correlation, we opted for an asymptotic Portmanteau Test, to address the heteroscedasticity assumption, we opted for the ARCH test. We then tested for normality using three different tests (Jarque-Bera test, Skewness-based test, Kurtosis-based test). To assess model stability, we used the help of an empirical fluctuation process (EFP), i.e. the OLS (Ordinary Least Squares) based CUSUM (Cumulative Sum) chart, which doubled as a model surveillance tool. We also opted for the Impulse Response Function, to determine the impact of a $1\sigma$ (or, one standard deviation) shock of precipitation and temperature on water level. Moreover, we carried out a Forecast Error Variance Decomposition (FEVD) technique to assess the extent to which the forecast error variance of individual variables can be attributed to external disturbances affecting other variables.
\subsection{Directional Dependence using Copula}
We have defined directional dependence along the lines of \cite{Kim2019}, i.e. the asymmetric dependence between random variables. The authors also suggested using bivariate copulas to determine the dependency structure between two random variables independently of any one-to-one continuous transformations of each variable. Let $C (u,v)$ denote a copula function, which defines the joint distribution of two random variables $(U, V)$ whose marginal distributions have a uniform distribution on $\left[0, 1\right]$. Let $C_u(v)$ denote the conditional distribution of $V$ given $U = u$ expressed as
\begin{equation}
    C_u\left(v\right)=P\left(V\le v\ |\ U=u\right)=\frac{\partial C\left(u,v\right)}{\partial u}
\end{equation}
The conditional expectation of $V$ given $U =u$ will be
\begin{equation}
    r_{U|V}\left(u\right)=E\left(V\ |\ U=u\right)=1-\int_{0}^{1}{C_u\left(v\right)\ dv}
\end{equation}
Hence, the copula directional dependence (CDD) of $V$ on $U$, or from $U$ to $V$, denoted by $\rho_{U\rightarrow V}^2$, can be interpreted as the proportion of variance of $V$ explained by the copula regression function $r_{U|V}(u)$. Similarly, the proportion of variance of $U$ explained by the copula regression function $r_{U|V}(v)$ can be measured using CDD of $U$ on $V$, or from $U$ to $V$ denoted by $\rho_{V\rightarrow U}^2$, where
\begin{equation}
    \rho_{U\rightarrow V}^2=\frac{Var\left[r_{U|V}\left(u\right)\right]}{Var\left(V\right)}=\frac{E\left[\left(r_{V|U}\left(U\right)-\frac{1}{2}\right)^2\right]}{\frac{1}{12}}=12E\left[\left(r_{V|U}\left(U\right)\right)^2\right]-3
\end{equation}
Similarly,
\begin{equation}
    \rho_{V\rightarrow U}^2=12E\left[\left(r_{U|V}\left(V\right)\right)^2\right]-3
\end{equation}
\textbf{Note:} The CDD defined above is a version of the Spearman’s Correlation Coefficient.\par  
Now, for estimating the CDD, determination of a parametric form of the copula regression function, $r_{U|V}\left(u\right)$, is necessary. Note that, both $U =F_X(X)$ and $V =F_Y(Y)$ are uniformly distributed and take their values in the unit interval $[0, 1]$. Since $r_{V|U}\left(u\right)$ is a conditional expectation of $V$ given $U = u$, both the response variable and predictor variable have bounded ranges. We have considered the Gaussian Copula Beta Regression for this purpose, in which, the cumulative distribution function, $F\left(v_t\ |\ u_t,\ \kappa_t\right)$ is used to transform the response variable $v_t$ into $w_t=F\left(v_t\ |\ u_t,\ \kappa_t\right)$, and subsequently used to obtain a standard normal variate $\epsilon_t$, using the inverse of the probability integral transform
\begin{equation}
    \Phi^{-1}\left(F\left(v_t\ |\ u_t,\ \kappa_t\right)\right)=\epsilon_t
\end{equation}
where $\Phi\left(.\right)$ denotes the cumulative distribution function of the standard normal distribution. It is this relationship between $v_t$ and $\epsilon_t$ that is used to formulate the sampling distribution and the likelihood function.
\section{Results} \label{sec4}
The final VAR model system was found to be:
\begin{align} \label{VAR}
    X_{1t} \nonumber &=-6.714-0.127\ X_{3\left(t-1\right)}-0.146\ X_{2\left(t-1\right)}+0.094\ X_{1\left(t-1\right)}+0.056 X_{3(t-2)}\\ \nonumber
    &+0.091 X_{2(t-2)}+0.011 X_{1(t-2)}-0.260 X_{3(t-3)}+0.039 X_{2(t-3)} \\ \nonumber
    &+0.291 X_{1(t-3)}-0.046 X_{3(t-4)}+0.430 X_{2(t-4)}+0.031 X_{1(t-4)} \\ \nonumber
X_{2t} &= 26.991-0.071\ X_{3\left(t-1\right)}-0.301X_{2\left(t-1\right)}+0.116\ X_{1\left(t-1\right)}-0.098\ X_{3\left(t-2\right)}\\ \nonumber &-0.182\ X_{2\left(t-2\right)}+0.027\ X_{1\left(t-2\right)}-0.018\ X_{3\left(t-3\right)}-0.195\ X_{2\left(t-3\right)}\\ \nonumber &-0.105\ X_{1\left(t-3\right)}+0.031X_{3\left(t-4\right)}+0.658\ X_{2\left(t-4\right)}+0.059X_{1\left(t-4\right)}\\ \nonumber
X_{3t} &=-7.485-0.021\ X_{3\left(t-1\right)}+0.473\ X_{2\left(t-1\right)}-0.019\ X_{1\left(t-1\right)}-0.022\ X_{3\left(t-2\right)}\\ \nonumber &-0.154\ X_{2\left(t-2\right)}+0.019\ X_{1\left(t-2\right)}+0.047\ X_{3\left(t-3\right)}+0.029\ X_{2\left(t-3\right)}\\ 
&-0.149\ X_{1\left(t-3\right)}+0.378\ X_{3\left(t-4\right)}+0.047\ X_{2\left(t-4\right)}-0.062\ X_{1\left(t-4\right)}
\end{align}
where, $X_{1t}$ denotes the water level at time $t$, $X_{2t}$ denotes the temperature at time $t$, and $X_{3t}$ denotes the precipitation level at time $t$. \par 
On testing the model residuals for serial autocorrelation, the asymptotic Portmanteau Test yielded a $p$-value of 0.1163 (> 0.05), indicating that no significant serial autocorrelation was observed at a $5\%$ level of significance.\par 
\begin{figure}[ht]
    \centering
    \includegraphics[width=1\linewidth]{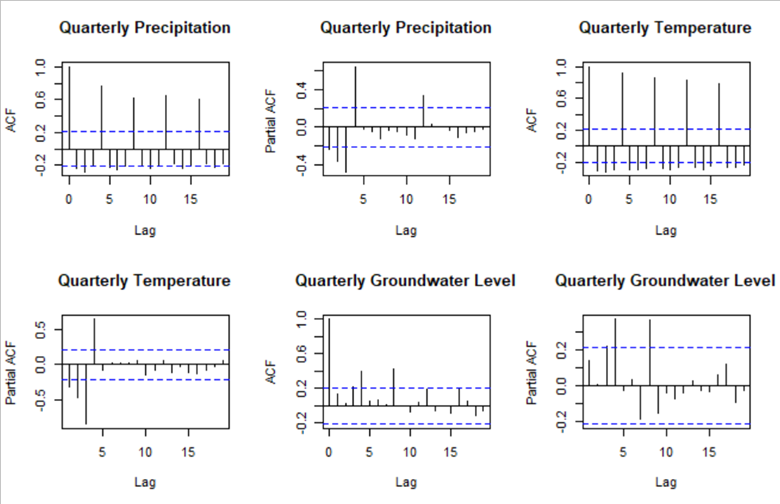}
    \caption{\textbf{(From left to right:)}ACF plot of quarterly precipitation, PACF plot of quarterly precipitation, ACF plot of quarterly temperature, PACF plot of quarterly temperature, ACF plot of quarterly groundwater level and PACF plot of quarterly groundwater level measured at Patyapura groundwater station/well from the first quarter of 2020 to the first quarter of 2021.\\ 
    \textbf{Data source:} India - Water Resource Information System}
    \label{fig:figure7_1}
\end{figure}
\begin{figure}[ht]
\centering
\begin{tabular}{ccc}
\includegraphics[width=0.465\textwidth]{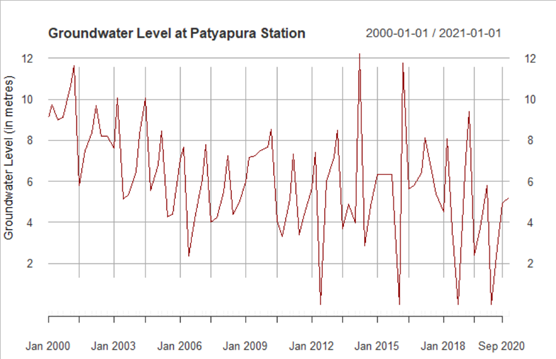} & \includegraphics[width=0.4\textwidth]{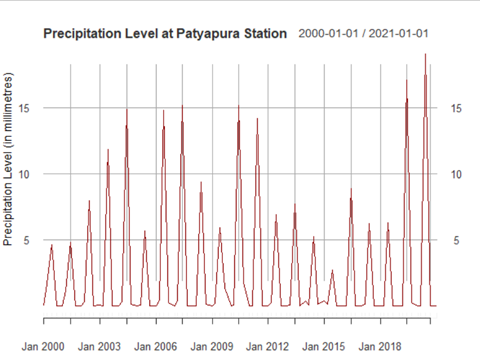} & \\
(a) & (b)& \\
\multicolumn{3}{c}{\includegraphics[width=0.4\textwidth]{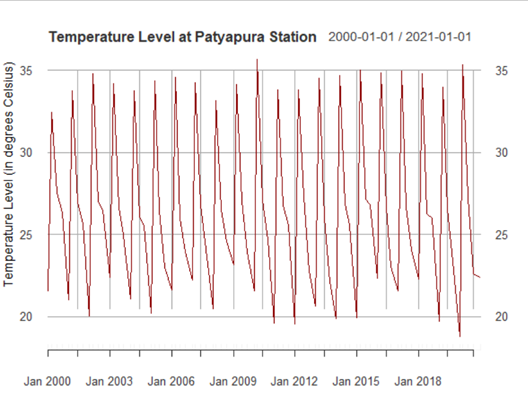}} \\
\multicolumn{3}{c}{(c)} \\
\end{tabular}
\caption{Run Charts of (a) quarterly groundwater level, (b) quarterly precipitation level, and (c) quarterly temperature level measured at Patyapura groundwater station/well from the first quarter of 2020 to the first quarter of 2021. \textbf{Data source:} India - Water Resource Information System}
\label{fig:figure7_2}
\end{figure}
\begin{figure}[ht]
\centering
\begin{tabular}{cc}
\includegraphics[width=0.5\textwidth]{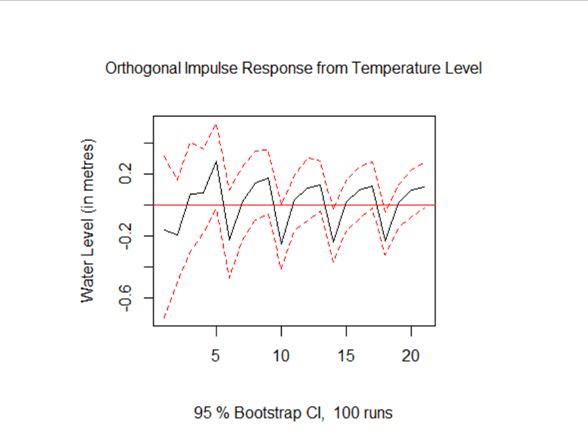} & \includegraphics[width=0.5\textwidth]{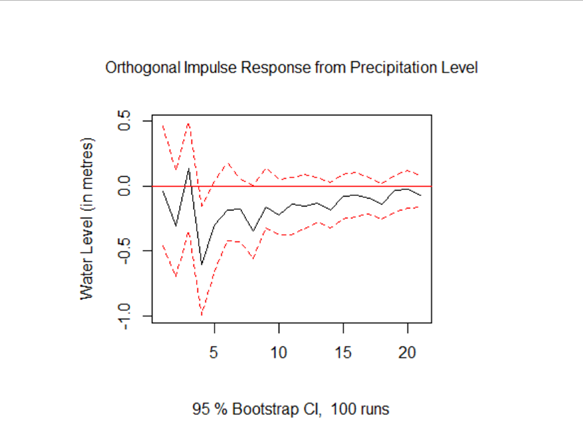} \\
(a) Precipitation Level & (b) Temperature Level \\
\end{tabular}
\caption{Impulse responses of precipitation and temperature levels on groundwater level at Patyapura Station/Well, Vadodara, Gujarat, India. \textbf{Data source:} India - Water Resource Information System}
\label{fig: figure7_3}
\end{figure}

The ARCH test on model residuals yielded a p-value of 0.7252 (> 0.05), indicating that no heteroscedasticity was observed in the sample, at a 5\% level of significance. Among the normality tests, the Skewness Test reported a p-value of 0.239 which fails to reject the null hypothesis that the residuals are normally distributed at 5\% level of significance. The plots for impulse response functions are provided in Fig. \ref{fig: figure7_3}, along with the run charts in Fig. \ref{fig:figure7_2} and the ACF and PACF plots of the individual time series of each variable in Fig. 4.1. \par 
\begin{figure}[ht]
    \centering
    \includegraphics[width=1\linewidth]{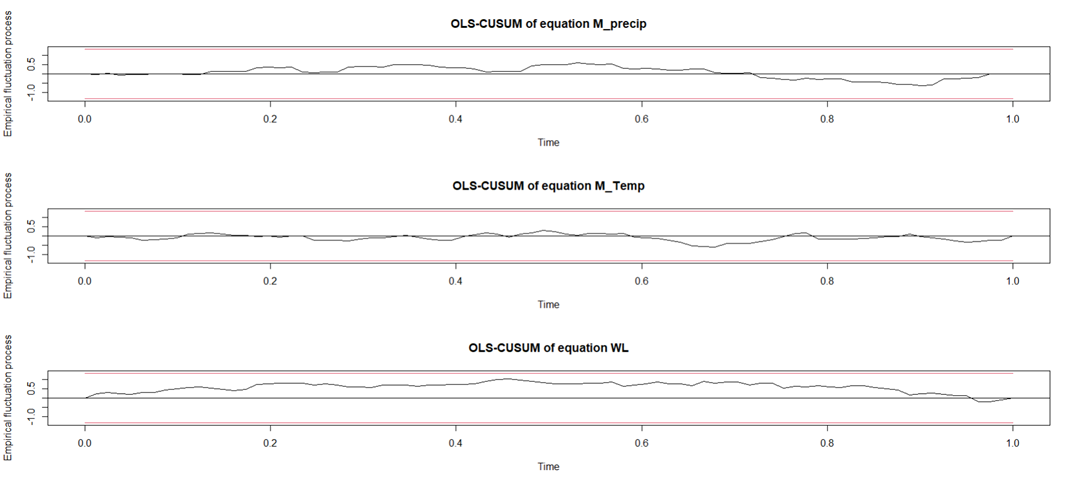}
    \caption{Empirical Fluctuation Process plot: OLS-CUSUM chart indicating structural changes in the residuals of the three models in the VAR system, described in Equation \ref{VAR}.}
    \label{fig:figure7_4}
\end{figure}
The primary takeaway from the Empirical Fluctuation Process (EFP) plot (Fig. \ref{fig:figure7_4}) is that it reinforces the notion that there has been no significant structural change observed with respect to time in any of the three variables taken into consideration at a 5\% level of significance. From the Impulse Response Functions (IRF) (Fig. \ref{fig: figure7_3}), we can see that in case of an initial one-sigma shock to the Precipitation level, the Groundwater level does stabilize over a period of 20 or so runs, but in case of an initial one-sigma shock to the Temperature level, we see that the Groundwater Level, does not converge, but maintains its seasonal stability. This may arise from the contemporaneous effect of Temperature on Groundwater level, and the seasonality inherent in both the variables. This is further supported by the Granger Causality Test on the model in which Temperature is the endogenous variable and Groundwater level and Precipitation level are the exogenous variables, that yielded a $p$-value of $0.00003916$, indicating that Temperature has a one-way impact on the variability of Groundwater level as well as Precipitation Level, at $5\%$ level of significance. Note that, the IRF assumes  contemporaneous association between exogenous and endogenous variables. \par 
\begin{figure}[ht]
    \centering
    \includegraphics[width=1\linewidth]{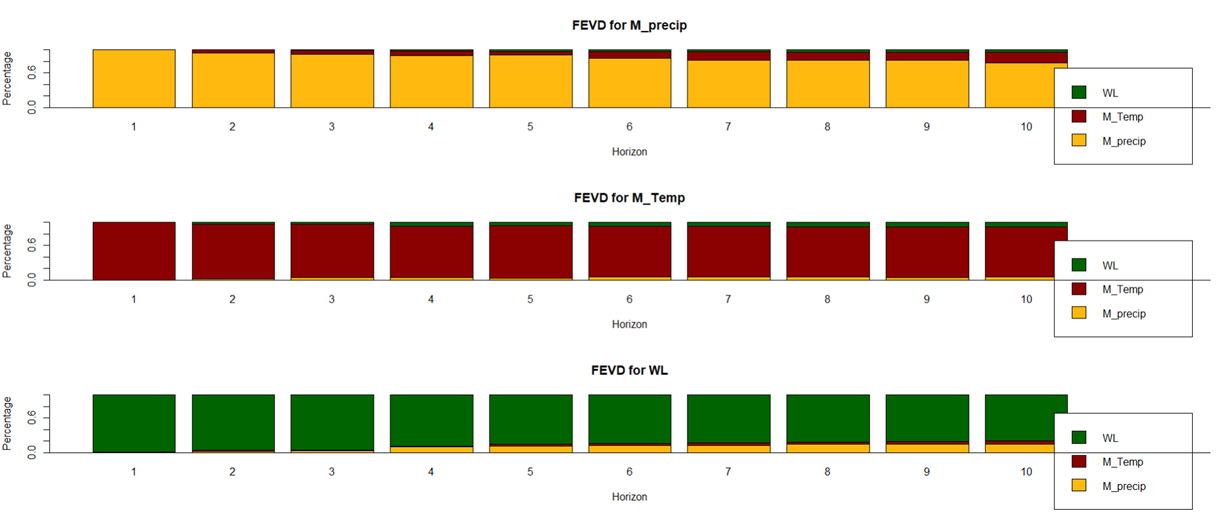}
    \caption{\textbf{(From top to bottom):} Forecast Error Variance Decomposition (FEVD) charts based on the three models in the VAR system described in Equation \ref{VAR}, indicating the amount of information contributed by groundwater level and temperature to the precipitation level, the amount of information contributed by precipitation and groundwater level to the temperature level, and the amount of information contributed by precipitation and temperature to the groundwater level in the autoregression.}
    \label{fig:figure7_5}
\end{figure}
The process of Forecast Error Variance Decomposition (FEVD) is a structural analysis technique that involves breaking down the variance of forecast errors into individual contributions from specific external shocks, as shown in Fig. \ref{fig:figure7_5}. It serves to reveal the extent to which a shock is significant in explaining variations in the variables of the model, and illustrates how the level of importance varies over time. Here, we can see that in 10 runs (quarters), Precipitation and Temperature contribute to almost $20\%$ of variability, increasing gradually since initial shock.\par
\begin{figure}[ht]
    \centering
    \includegraphics[angle=90,origin=c,width=0.9\linewidth]{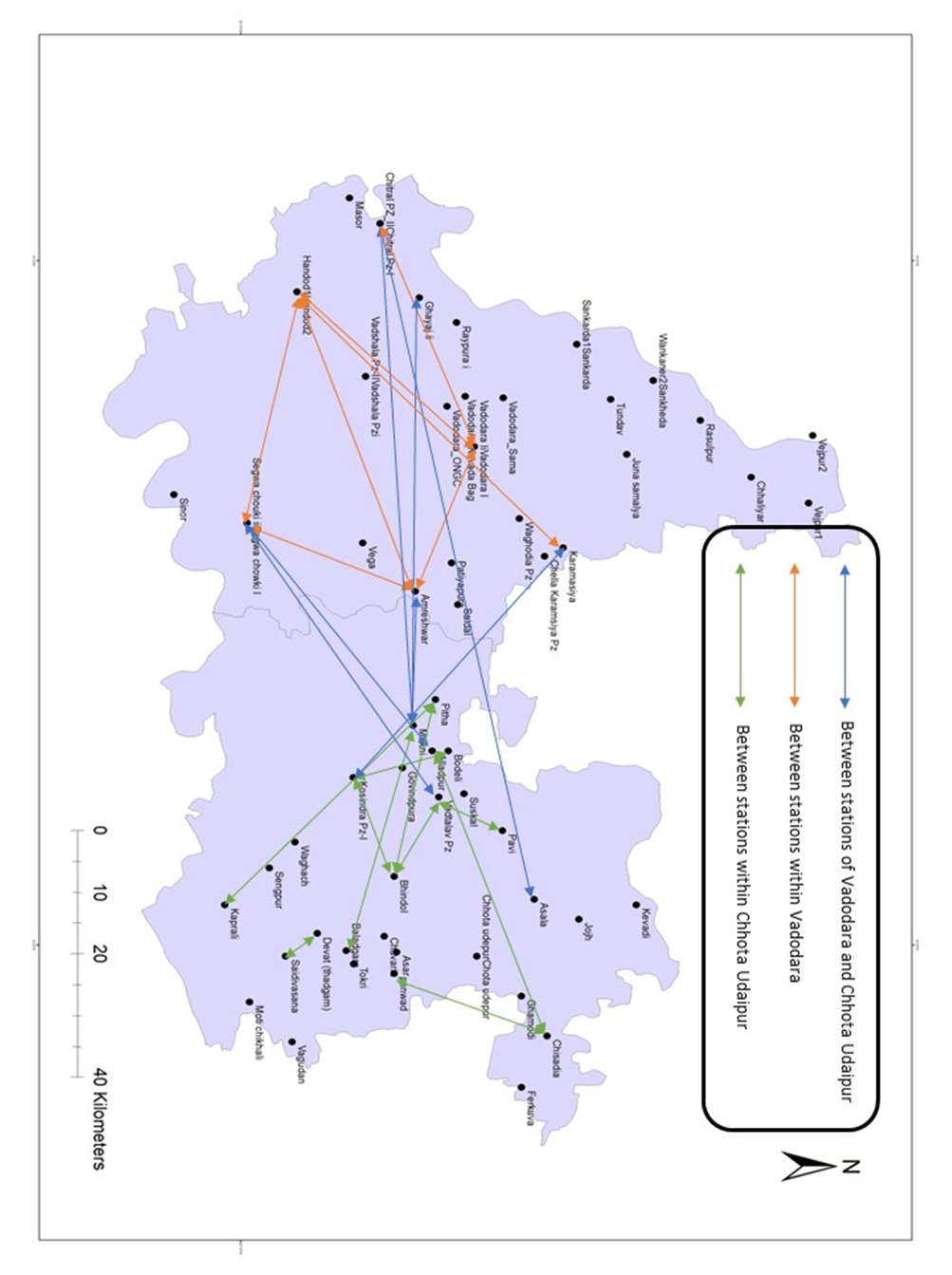}
    \caption{Map depicting the dependency network of groundwater stations/wells in the districts of Vadodara and Chhota Udaipur, in Gujarat, India, having Copula Directional Dependency (CDD) higher than 0.95, for selected pairs of stations in the districts of Vadodara and Chhota Udaipur. \textbf{Data source:} India - Water Resource Information System}
    \label{fig:figure7_6}
\end{figure}

 
The GCBR-based CDD measure was obtained for all pairs of stations and Table \ref{tab:table7_1}. enlists the pairs having maximum dependence between each other. The same has been illustrated by a map of Vadodara district and the Chhota Udaipur (Udepur) districts in Fig. \ref{fig:figure7_6}. Additionally, in Fig. \ref{fig:figure7_6}, we have attempted to map dependencies within and between districts. Note that, only pairs in which the CDD was greater than 0.95 were considered for interpretation.\par 
\begin{table}[htbp]
  \centering
  \caption{List of groundwater stations between which copula directional dependence was found to be greater than 0.95}
    \begin{tabular}{|c|c|c|c|}
    \hline
    \multirow{1}{*}{\textbf{Station 1 ($U$)}} & \multirow{1}{*}{\textbf{Station 2 ($V$)}} & \multirow{1}{*}{\textbf{$\rho_{U \rightarrow V}$}} & \multirow{1}{*}{\textbf{$\rho_{V \rightarrow U}$}} \\
      \hline
    Alladpur & Chisadia & 1     & 0.94683 \\
    \hline
    Alladpur & Segwa Chowki I & 0.96587 & 0.9674 \\
    \hline
    Amreshwar & Handod I & 0.96479 & 0.96443 \\
    \hline
    Amreshwar & Makni & 0.97426 & 0.974 \\
    \hline
    Amreshwar & Segwa Chouki II & 0.98572 & 0.98474 \\
    \hline
    Amreshwar & Vadodara II & 0.96374 & 0.96193 \\
    \hline
    Asala & Chitral PZ II & 0.97568 & 0.9752 \\
    \hline
    Baladgam & Makni & 0.95694 & 0.95908 \\
    \hline
    Bhindol & Kosindra Pz I & 0.99311 & 0.9933 \\
    \hline
    Bhindol & Pitha & 0.96079 & 0.96143 \\
    \hline
    Bhindol & Vadtalav PZ & 0.97362 & 0.97328 \\
    \hline
    Bodeli & Kosindra PZ I & 0.96573 & 0.96685 \\
    \hline
    Chisadia & Panwad & 0.9631 & 0.96486 \\
    \hline
    Chitral PZ II & Makni & 0.99218 & 0.99237 \\
    \hline
    Chitral PZ II & Vadodara I & 0.96172 & 0.96216 \\
    \hline
    Devat (Thadgam) & Saidivasana & 0.95406 & 0.96142 \\
    \hline
    Ghayaj II & Makni & 0.99612 & 0.99629 \\
    \hline
    Handod I & Karamasiya & 0.95198 & 0.95206 \\
    \hline
    Handod I & Segwa Chouki II & 0.98175 & 0.98154 \\
    \hline
    Handod I & Vadodara I & 0.95313 & 0.95351 \\
    \hline
    Kaprali & Pitha & 0.97459 & 0.97362 \\
    \hline
    Karamasiya & Kosindra PZ I & 0.95942 & 0.95911 \\
    \hline
    Pavi  & Vadtalav PZ & 0.97843 & 0.9787 \\
    \hline
    Segwa chouki II & Vadtalav PZ & 0.97647 & 0.97773 \\
    \hline
    \end{tabular}%
  \label{tab:table7_1}%
\end{table}%

\begin{figure}[t!]
\centering
\begin{tabular}{c c}
\includegraphics[width=0.5\textwidth]{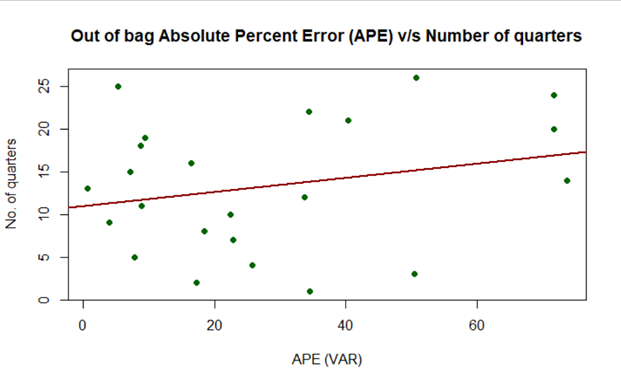} & \includegraphics[width=0.5\textwidth]{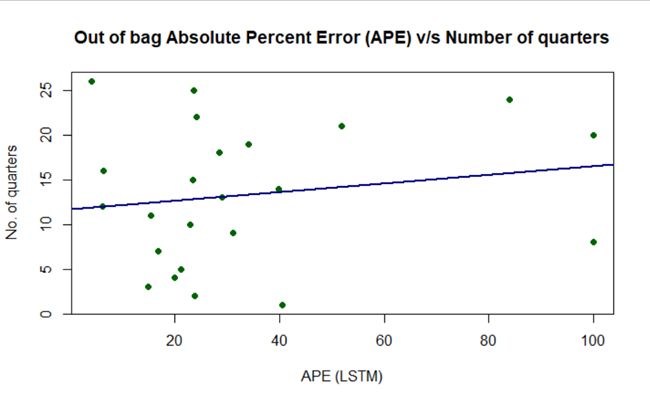} \\
(a) & (b) \\
\end{tabular}
\caption{Out of Bag Error Plots indicating the relationship between forecast error and the time between the last training data point and the last forecasted point, measured in quarters, used in hte calculation of Model Shelf-Life of (a) Vector Auto Regression model and (b) Long Short-Term Memory - Recurrent Neural Network}
\label{fig: figure7_7}
\end{figure}
From the fitted model, we observed that the expected shelf life of the VAR model was 11 time points or 11 quarter-years, or 33 months, i.e. 2 years 9 months. It indicates that at the current stage, it is expected that the model shall be good for (i.e. shall not provide APE greater than $5\%$) 11 quarter-years. Any model that provides higher shelf life can be considered a more desirable model. The LSTM model generated a shelf life of 12 quarter-years. Plot of the regression line on which Model Shelf-Life is calculated is provided in Fig. \ref{fig: figure7_7}.
\section{Conclusions} \label{sec5}
From the above results, we have observed that across all the groundwater stations, across time as well, the relationship between groundwater level, precipitation and temperature have been consistent and their dependence structures are more or less stable. Speaking across stations, although it has been observed that some stations have high dependencies with multiple other stations, there were a few stations namely, Vadtalav, Makni, Amreshwar, Segwa Chowki, Vadodara and Handod that have high dependencies with most other stations, implying that these are spots of dependency concentration. One of the reasons it might prove useful is that instead of having to monitor all stations, a thorough investigation may be taken sporadically based on changes in these stations only. Additionally, Amreshwar, Segwa Chowki and Makni stations are spots having maximum dependencies with stations outside their district. Much of it can be attributed to the location of these stations; from a cursory glance at Fig. \ref{fig:figure7_6}, one can observe that these stations are closest to the district boundaries. Almost all of the dependencies that have been obtained have been bi-directional. However, additional historical data may provide more contrasting results. The dependencies having largest spatial distances may be attributed to common aquifers or similar elevation levels. \par    
Turning our focus towards the model shelf-life, judging the VAR model based solely on obtained shelf-life based on APE may not be in the best interest. The LSTM is conventionally a model aimed for forecasting (if at all intended) univariate time series data (loosely said; it works even for at most multiple, but not necessarily uniform, replications of the same variable at multiple time points, intervals between said time points, not necessarily equal). The case under study was that of a multivariate time series data, which means that the VAR model has an edge due to it being fed more information in the form of the precipitation and the temperature data, which the LSTM model is not exposed to. One may choose a more balanced comparison by either considering a simple ARIMA model on the groundwater level values and comparing its shelf life to that of the LSTM, or by comparing the VAR model to some multivariate GRU-RNN or some Temporal CNN. Alternatively, one may choose to use some RNN in which the LSTM architecture is modified to accommodate additional features, longitudinally.
\section{Future Scope and Open Research Areas} \label{sec6}
Judging from most of the papers cited, one can see that more recent papers show a clear, albeit slight, shift in model usage from statistical to numerical/mathematical, especially in case of GWL modelling. A plausible explanation to this shift may be, the abundance of known parameters with which to model. One might argue theoretically, from a semi-parametric standpoint, that no countably finite number of parameters is sufficient to model non-experimental data deterministically. However, considering too many parameters might not always result in a desirable situation for most researchers, as the interpretation may become too complex, giving rise to the need of modern model regularization techniques, which may be of use to researchers. \par 
It has been observed, though, that statistical models are being used heavily in case of Groundwater Recharge Zone Mapping, and other classification models. Ever since Neural Networks have become computationally cheap and easily accessible via various software, model comparisons, or more accurately, machine-learning-pipeline-architecture comparison has become commonplace. An upward spike in the popularity of hybrid models, especially in case of GWL modelling, solidifies the notion that collective effort is being directed towards forecasting GWL using decreasing number of parameters. However, we have also come across papers, although very few, that have extensively disregarded, or in some cases, have taken arguably flawed decisions related to model validation (ref. Appendix \ref{secA1}). Real-time monitoring systems are a rarity in this field, as we have observed. There are dynamic models that exist in this sphere, but a symbiotic effort to lay down systemized protocols for usage of their respective models, or to provide inputs for model upgradation has become essential nowadays.
\subsection{Future Scope}
In their respective review articles, \cite{Rajaee2019, Machiwal2018} have discussed key points which are yet to be addressed, like, focusing on the synthesis and verification aspects of time series modelling, which could potentially be achieved using stochastic modelling of groundwater data, incorporating characteristics like homogeneity, stationarity, persistence and periodicity, focusing on Bayesian kriging methods as opposed to mainstream kriging, developing protocols for monitoring of groundwater quality for comparative appraisal of the vulnerability degree of aquifers, developing robust methodologies to deal with the groundwater quality index (GWQI) related studies, or to some extent, developing a generalized version of it to enable unanimous comparison, overcoming the inevitable trade-off involved when dealing with hybrids of AI models with conceptual-numerical models, among others. \par 
With the kind of easement in terms of access to GIS technologies that we have today, the scope of research is limitless. However, there remains a dearth of tools that can be accessed by researchers, due to excessive reliance on software applications like ASTER-DEM, ArcGIS, etc., wherein researchers are compelled to restrict themselves to whichever methods or models are available within said applications, and hence, the potential for a varied, out-of-the-box statistical regimen might not be fully realized. This certainly does not imply the complete absence of software applications, or programming languages, that enable researchers to analyse geospatial data in a customized manner, although that may require a certain degree of expertise in programming.\\

\bmhead{Acknowledgements}
We would like to thank the esteemed reviewers and the editorial board at the International Journal of System Assurance and Engineering and Management for their valuable feedback and their suggestions for improving the quality of the article.

\section*{Declarations}
The authors hereby declare that no funding was received for the research, analysis and/or writing of this paper, by any financial or non-financial institution.\par  
The authors certify that they have no affiliations with or involvement in any organization or entity with any financial interest (such as honoraria; educational grants; participation in speakers’ bureaus; membership, employment, consultancies, stock ownership, or other equity interest; and expert testimony or patent-licensing arrangements), or non-financial interest (such as personal or professional relationships, affiliations, knowledge or beliefs) in the subject matter or materials discussed in this manuscript. \par 
All data and materials including software and programs built, support published claims and comply with field standards. All data used can be easily procured (and simulated, in some cases) from the GitHub repository: \url{https://github.com/dasagniva/groundwater}. \par 
During the preparation of this work the authors used no AI tool to write, analyse the data related to, or embellish the research article in any form. The authors themselves reviewed and edited the content as needed and take full responsibility for the content of the publication.
\newpage    
\begin{appendices}
\section{Tabulated review of some spotlight research articles}\label{secA1}
\begin{longtable}{p{3em} p{4.28em} p{8.665em} p{6.5em} p{4.78em} p{8.665em}} \caption{List of reviewed research articles corresponding to their statistical goals or objectives, a summary of their methodologies, methods of model validation employed, software used, name of journal in which it was published, country of the journal where published and variables used in the study}  
  \label{tab:appendix_table}%
  \\
    \hline
    \textbf{Article} & \textbf{Objective} & \textbf{Methodology} & \textbf{Validation} & \textbf{Software Used} & \textbf{Variables used} \\
    \hline
    \cite{Jaafarzadeh2021} & GWR Zone Classification & Classifier Ensemble: Maximum Entropy Model, Frequency Ratio Model & 70-30 Hold Out CV: Receiver Operation Characteristic (ROC) approach & {ArcGIS}  & Elevation, Slope Percentage, Slope Aspect, Profile Curvature, Plan Curvature, Drainage Density, Distance from Rivers, TWI, Stream Power Index, Rainfall, Lithology, Fault Distance, Land Use, NDVI, Soil Texture \\
    \hline
    \cite{Azma2021} & GWR Potential map & Multiple Regression, Principal Component Regression & 70-30 Hold Out CV: Coefficient of Determination (R-square) and Root Mean Square Error (RMSE) & {SAGA GIS, ArcMap 10.6, Statistica 8.0} & Clay, CACO3, Electrical Conductivity, Elevation, Slope, TWI, LS factor, Curvature, Flow Accumulation, Flow Direction \\
    \hline
    \cite{Bahir2021} & GW Quality Evaluation & Chemical Analysis, Piper Diagram, Correlation Analysis, Simple Linear Regression` & NA    & {MS Excel} & Groundwater depth, Electrical Conductivity, Temperature, pH, quantities of various chemicals and minerals \\
    \hline
    \cite{Amiri2021} & GW Quality Evaluation & Principal Component Analysis, Cluster Analysis & NA    & {MS Excel} & pH, Electrical Conductivity, Dissolved Oxygen, Total Organic Carbon, Various chemical quantities, Biological Oxygen Demand, Chemical Oxygen Demand, Total Dissolved Solids, Oil and Grease \\
    \hline
    \cite{carrion2021geometric} & Aquifer Management & Geometric Model: Digital Terrain Model, Geological Mapping & {NA} & {GeoModeller, IPI2win, ArcGIS, MS Excel} & Vertical Electrical Surroundings: Resistivity, Thickness, Depth; Project Coordinates, Geological Map: Lithology, Volume; Aquifer data: Static Level, Average porosity of gravel/sand layer, Darcy's Permeability, Well-Extraction Flow \\
    \hline
    \cite{Ouarani2021} & GW Quality Evaluation and GWR Modelling & Chemical Analysis, Piper Diagram, Cluster Analysis (Hierarchical, Agglomerative, using Ward's method), Factor Analysis (using principal factor extraction, with Varimax rotation) & NA & {IBM SPSS v26.0} & pH, Temperature, Electrical Conductivity, Total Dissolved Solids, Natural Parameters: Na, Cl, NO3, HCO3, Ca, SO4, K and Mg; Ionic Balance;  Anthropogenic impact represented by NO3; Processes controlling HCO3 concentration in groundwater \\
    \hline
    \cite{Khedri2020} & GW Level Estimation & ANN, Fuzzy Logic, Adaptive Neuro Fuzzy Inference System, Least Squares Support Vector Machine, Group Method of Data Handling & Hold-out CV (120 intial time points in training dataset; subsequent points in testing dataset): R-square, RMSE, Mean Absolute Error (MAE), Nash-Sutcliffe Efficiency (NSE) & Mathworks MATLAB & Monthly data: GW Level, Total Precipitation, Total Evapotranspiration \\
    \hline
    \cite{Yariyan2020} & GWR Potential map & To detect multicollinearity: Variance Inflation Factor (VIF) and tolerance; Regression Models used: Frequency Ratio (FR) model, Evidential Belief Function (EBF) model; Hybridizations: Radial Basis Function (RBF) NN, Index of Entropy (IoE) model, Fuzzy Art Mapping (FAM) NN & 70-30 Holdout CV; ROC approach: Area Under Curve (AUC), Positive Predictive Value (PPV), Negative Predictive Value (NPV), Sensitivity, Specificity, Accuracy; coupled with Friedman's Test & ArcGIS v10.4, SAGA GIS v3.3 & Topographic: Altitude, Slope, Slope Aspect, Curvature, Topographic Ruggedness Index (TRI), TWI; Earthly: Lithology, Soil, Distance to Fault, Distance to Stream; Hydrological: Rainfall, Drainage Density, Flow Direction; Landsar-8 OLI Imaginary: Land Use \\
    \hline
    \cite{Lian2020} & {Rainfall Simulation} & Bagged Classification Tree (Random Forest) hybridized with ANN (single hidden layer having tan-sigmoid function, output layer having linear transfer function, backpropagated using Levenberg-Marquadt algorithm) compared opposite a Multivariate Non-Homogeneous Hidden Markov Model with coefficients estimated using Maximum Likelihood {Expectation Maximization (EM) algorithm used for computation} Method, optimum number of hidden states evaluated using BIC criterion & Probability of Detection (POD), False Alarm Rate (FAR), Heide Skill Score (HSS), Kolmogorov-Smirnov Test (for goodness of fit), Mann-Whitney U-test, For normality: Anderson-Darling, Jarque-Bera, Lilliefors tests & {Mathworks MATLAB} & Daily rainfall from 1975 - 2012, Circulation variables: Geopotential and Wind Component; Temperature, Radiation, Moisture variables: specific humidity (Total of 26 variables considered; others not mentioned) \\
    \hline
    \cite{Fung2020} & Drought Prediction through prediction of Standardized Precipitation Evapotranspiration Index (SPEI) & Support Vector Machines: Fuzzy SVR (generated fuzzy membership values as additional inputs and trained with SVR) and Boosting SVR (Boosted Wavelet Transformation over SPEI and trained in SVR); SPEI calculation involved Parameter Weighted Moments (PWM's) using L-moment method & MAE, RMSE, Mean Bias Error (MBE), R-squared & Mathworks MATLAB & Month-wise data collected: Precipitation, Temperature, Rainfall, Latitude -Longitude, Heat Index, Number of Sun Hours, Number of days in the month \\
    \hline
    \cite{Natarajan2020} & GW Level Forecasting & Comparison between: ANN: Feed-forward with number of hidden layers optmized by trial-and-error, backpropagation using Gradient Descent Optimization (GDO); Genetic Programming (GP) Simulation; Extreme Learning Machine: Single Layer Feed-forward NN model, sigmoid approximation function; SVM and its hybrid varieties: SVM-QPSO (Support Vector Machine - quantum-behaved - Particle SWARM Optimizer), SVM-RBF (Support Vector Machine - Radial Basis Function) & RMSE, R-squared, Pearson's Correlation Coefficient, MAE, Mean Absolute Percent Error (MAPE) & {Undisclosed; Graphs generated using MS Excel (visually deduced)} & Monthly Groundwater Levels and Rainfall data of six different locations from 2007 - 2012 \\
    \hline
    \cite{Banadkooki2020} & GW Level Estimation & A Multi Layer Perceptron (MLP) hybridized with the Whale Algorithm (WA): Single hidden layer, parameters optimized using evolutionary algorithms, sigmoid transfer function as activation; Radial Basis Function (RBF) Neural Network hybridized with the Whale Algorithm: Single hidden layer (RBF-based), sigmoid transfer function as activation; Genetic Programming (GP)Simulation Model; Partial and Mutual Autocorrelation Analysis to fix time-delays and generate various scenarios & R-Squared, MAE, Nash-Sutcliffe Efficiency Coefficient (NSE), RMSE-observations Standard-deviation Ratio (RSR) & Undisclosed & Collected at lags of 3, 6 and 9 months: Temperature, Rainfall; GWL; Input data modified to various combinations of the lagged variables \\
    \hline
    \cite{Miraki2019} & GWR Potential map & Classifier Ensemble: Random Subspace based Random Forest (RS-RF); Feature selection using LSSVM; Comparison of RS-RF with Random Forest (RF), Logistic Regression, Naïve Bayes & For LSSVM: 10-fold CV; For RS-RF: ROC approach (AUC considered), Success and Prediction rate curves, Kappa Index; For comparison with classical models: Friedman Test, Wilcoxon Signed Rank Test & {ENVI 5.1, DEM (Software undisclosed), No computational software mentioned explicitly} & Slope, Aspect, Elevation, Curvature, Stream Power Index (SPI), TWI, Rainfall, Lithology, Land Use, NDVI, Fault Density, River Density \\
    \hline
    \cite{Mohammed2022} & GW Quality Evaluation & Vanilla LSTM calibrated using process-based hydrological model, namely, the Generalized Longitudinal Lateral-Veritcal Hydrodynamic Transport (GLLVHT) model, wherein output of the process-based model based on multi – variable input data, used to validate and train LSTM for a univariate time series. & Model performance was judged based on R-squared, Nash-Sutcliffe Model Efficiency (NS) metric and Mean Squared Error (MSE) based on a 60:40 train – test Holdout split & Process-Based Model implemented using Generalized Environmental Modelling System for Surface Waters (GEMSS) software (GEMSS-HDM module), LSTM programmed in Python using Keras and Tensorflow packages & Dynamic: Weather variables (hourly data collected over 7 months: Air temperature, dew point temperature, net solar radiation, cloud cover, barometric pressure, wind direction, wind speed, relative humidity ; Monthly data collected over 7 months: E.coli concentration, coliform bacterial concentration, heavy metal concentration, Inflow and discharge, Outflow and withdrawals, Observed FIB (fecal indicator bacteria) inflows, Heavy Metals Inflows, Outflows and withdrawals); Spatial: Depth, Sechi depth, Vegetative and topographic cover,  Shape and Bathymetry GIS \\
    \hline
\end{longtable}%
\newpage




\end{appendices}

\bibliography{sn-bibliography-1}%

\end{document}